# Emergent Insight of the Cyber Security Management for Saudi Arabian Universities: A Content Analysis

**Hamzah Masmali and Shah J. Miah**

Newcastle Business School, University of Newcastle, NSW, Australia

Email: shah.miah@newcastle.edu.au

**Abstract**

While cyber security has become a prominent concept of emerging information governance, the Kingdom of Saudi Arabia has been dealing with severe threats to individual and organizational IT systems for a long time. These risks have recently permeated into educational institutions, thereby undermining the confidentiality of information as well as the delivery of education. Recent research has identified various causes and possible solutions to the problem. However, most scholars have considered a reductionist approach, in which the ability of computer configurations to prevent unwanted intrusions is evaluated by breaking them down to their constituent parts. This method is inadequate at studying complex adaptive systems. Therefore, the proposed project is designed to utilize a holistic stance to assess the cybersecurity management and policies in Saudi Arabian universities. Qualitative research, entailing a thorough critical review of ten public universities, will be utilized to investigate the subject matter. The subsequent recommendations can be adopted to enhance the security of IT systems, not only in institutional settings but also in any other environment in which such structures are used.

**Keywords:** Cyber security, public universities, cybercrime, educational institutes

# 1 Introduction

Information security is among the top concerns for virtually all organizations across the globe. Firms intending to protect their data have to identify all gaps that may increase their vulnerability, as well as procedures of averting the potential of misusing critical records (Dehlawi & Abokhodair, 2013). More importantly, the mere development and implementation of seemingly adequate data security policies and guidelines to safeguard themselves from potential risks are not enough. Organizations have to continuously create, maintain, and improve their security systems to mitigate both internal and external threats (Reid & Van Niekerk, 2014). Nevertheless, the proliferation of portable storage, wireless, and other computing technologies has significantly increased the cybersecurity risks that many companies face.

Many universities in today's world admit that they are struggling to deal with issues related to cybercrime, particularly on access to innovative ways of managing and protecting data (Cheung, Cohen, Lo, & Elia, 2011). The public-funded universities are particularly financially vulnerable compared to private ones. Consequently, such economically challenged institutions are open to collaborative efforts. Besides, education facilities have limited or no control over the websites since they are accessible to students and teachers who can even login using their devices (Ramim & Levy, 2006). Furthermore, schools receive new students every year while others complete their educations, therefore, building a tidal wave of cyber insecurities. To that effect, learning institutions should come up with a holistic move towards striking a balance and ensuring data is protected without prohibiting or blocking students from accessing the platforms. While the region's education system is striving to provide world-class quality for learners' and teachers' experience, they seem to lag in Information Technology (IT) security (Rezgui & Marks, 2008). It is vital that they adequately invest in defending cybercrime incidences by adopting the right technologies.

The exponential growth, progress, development and complex possibilities that the cyberspace provides for all sectors have been the cause for concern in recent days. Cybercrimes are on the rise and breach of cybersecurity is a threat that needs to be handled ethically, legally, scientifically and quickly since our lives depend on cyberspace (De Bruijn & Janssen, 2017). Schools, colleges and universities have started to capitalize on the multiple options provided by cyberspace and are using it extensively for both academic works as well as all kinds of official work. With so much sensitive data being handled, authorities must focus on Cyber Security so that the data remains safe and out of reach of wrong hands. This study looks into the cybersecurity management and policies of 10 selected universities in Saudi Arabia.

The research focuses on the policies and approaches of the universities for maintaining the cybersecurity level. The cybersecurity awareness is important for the universities for protecting the information assets and maintaining the privacy of the management and students. The knowledge and application of the global policies and regulations for cybersecurity can be useful for the university managements to overcome the impact of the hacking, malware attack, phishing of the data and misuse of the cloud storage. In addition to this, information and knowledge of the multi-layer cybersecurity approach is the best way to protect the data and systematically manage the information. The research will make emphasis on the combination of firewalls, software and variety of tools that will help the universities to combat the situation of a cyber attack or consulting the experts for managing the digital attack on the university data (Pandey, and Misra, 2016). There are different types of AI tools that have been used to overcome the malicious cyber crimes. The knowledge and understanding of the cyber policies and approaches will help the universities of Saudi Arabia to improve the level of security.

**2 Background of the study**

**2.1 Cybersecurity**

Cybersecurity is the organisation and selection of tools, procedures and mechanisms used to secure sensitive information in cyberspace from unauthorized and criminals who might damage, misalign or harm organisations (Craigen, Diakun-Thibault, & Purse, 2014). This indicates that only authorised people can have access to sensitive information, such as software, hardware, data, and network in cyberspace. This is a clear definition relating to cybersecurity management in the education sector. There is a serious need to examine the potential cybersecurity risks that organisations face, especially regarding confidential information (Slusky & Partow-Navid, 2012). Present-day's cyber configurations go beyond the hardware and software components. They also include systemic economic, social and political aspects that are so interconnected that it has become virtually impossible to isolate the human element from the IT systems (Benson, McAlaney, & Frumkin, 2019). Although existing literature provides a wealth of knowledge on the social facet's influence on cyber operations, it does little to explain its relationship to cybersecurity.

According to (Alzahrani & Alomar, 2016), cybersecurity is the practice of defending computer, servers, mobile and electronic devices as well as networks from malicious attack. This type of term is known as information technology security and applied for a variety of context from the business. There are different types of cybersecurity functions are used for managing cybersecurity. This includes network security, application security, information security and operational security. These processes are having a significant impact on the promotion of cybersecurity and influence the approach of businesses and organizations. However, there are different types of threats that are influencing the approach of the management to maintain the effectiveness in cybersecurity. The education, medical and public entities are facing the various issues related to cybersecurity that influencing the data management and increasing the threats. The types of threats involve the cybercrime that includes targeting the system for financial gain. The second form of threats is cyber-attacks that are politically motivated for the gathering of the information for personal benefits. The third is cyber terrorism that is intended to undermine the electronic system to increase the fear in internet users.

**2.3 ICT Application and security issue**

The field of information and communication technology (ICT) has evolved with time. It has significantly increased any country's responsibility concerning citizens' security, in addition to having both peaceful and non-peaceful uses (Yeniyurt, Wu, Kim, & Cavusgil, 2019). This technology helps provide easy and secure ways of reaching valid information. In contrast, it can also lead to identity theft or access to personal information without permission, i.e. cyber-attacks.

Cybersecurity is the term used for the security of data. While this field affects multiple disciplines, the most significant sector affected by ICT is education. In the current scenario, the universities are using the ICT tools for managing the educational activities that help to maintain the communication 24*7. The online communication and sharing of the information are allowing the students and professors to get knowledge of process and issues. The implementation of ICT in education is increasing the value to teach and learning by enhancing the effectiveness by utilizing the videos and audio files. People connected to the internet using different devices such as computers, tablets and smart phones (De Bruijn & Janssen, 2017). The networking system of universities involves the administrative, management and educational data. The protection of these kinds of data is the major priority of the universities as it can harm the trust and reputation of the organization. The universities are become more concerned about cyber issues and using the different tools and technologies to overcome the threats.

The issues with ICT related to security are involving the privacy means, lack of understanding of the utilization of the tools and applications for connecting with the portal of the university. The third-party vendor services and unauthorised access to the data is also a big security threat for universities that are using the ICT services. In addition to this, the lack of knowledge related to ethical and legal policies for managing the data and security level is influencing the utilization of the technology. As per the views of the Ullah et al., (2019), the Distributed Denial of Services (DDoS) attacks are a most common type of cyber issues in ICT that affecting the level of education venue. The motive of such attack is to disruption to the institute's network, system or data centre that harms the productivity and system approach of the university. The target of such attack is poorly managed and protected portals of the universities. Another issue related to security for ICT is data theft. This affects the level of education by stealing the data of the students and staff such as name, mobile numbers, address, bank details and email. The hackers are selling the information to the third party or used as a bargaining tool and extort money.

The Internet, which is the flagship product of ICT, is a strong tool for cultural exchange and a source of wealth for human civilisation. The Internet offers people freedom never previously acquired a possibility of escape, but also of exchanges. While the relationship between different peoples through the network has its advantages, it is also dangerous and complex. The term Cybercrime was born at the end of the 90s, with the explosion of exchanges via 'the net'. This period was marked by increasingly frequent infringements on the Internet, such as violations of the rights to privacy or confidentiality. The arrival of the Internet, from the top debit, has led to the emergence of a new criminality category, cybercrime. Therefore, it was necessary to set up legislation to adapt to this new type of crime (Marcum & Higgins, 2019).

In the current scenario, major universities are offering the course through an online platform that is helping in engaging the students and staff members. The online classes and workshops help provide flexible learning and maintain communication among professors and students. The universities have developed websites that offering information related to the course and providing the study material. The students can log in into the portal through personal email id. This kind of facility is having a framework for maintaining the security as personal information of students and university management may not be used for personal benefits. Apart from this, the issues related to cyber hacking and data manipulation harming the reputation of the universities. The challenges related to cybersecurity issues that universities are facing in the current scenario involve the regulatory compliance, third-party vendor services, lack of expertise for protecting the data and open culture (Tweneboah-Koduah, Skouby, and Tadayoni, 2017). These issues are affecting the approach and work of universities in Saudi Arabia. University management needs to improve the security level to maintain the effectiveness and trust of the staff and students. The management is looking for improvement in cybersecurity by applying the new technology and verification codes that help in protecting the unauthorised access and hacking.

**Research questions**

1. What are the major cyber threats for the universities of Saudi Arabia?

2. Are the universities able to handle the threats?

3. What are the policies that could be useful for the universities to overcome the threats of cybersecurity?

This study empirically investigates and appraises the present state of cybersecurity management and policies of ten public universities in Saudi Arabian, with a specific focus on the potential cybersecurity breaches. It is important to have certain supporting objectives to complete the assessment:

- The first objective is identifying the risk and threat caused by cybercriminals against the universities in Saudi Arabia. There is a significant lack of risk and threat when it comes to cybersecurity for higher education regarding the importance and scale of a large volume of sensitive data.
- Second objective by examining whether the universities are ready to handle the threats and risks they face. Preparation must be determined by the level of cybersecurity management in the universities.
- Third objective by suggesting recommended policies to enhance the management of cybersecurity in the universities. The process of selecting and identifying specific recommendations will be driven by providing evidence including a recommendation in the context of research.

The implementation of cybersecurity measures requires large investment for software and firewalls. In addition to this, cultural issues like bring your own devices to increase the challenges for the universities to develop a secure wider network. By providing the basic training to the management and students for using the network and access of the personal portal can help secure the data. The management can provide a simple handbook for policies and tips for practising good cybersecurity hygiene. This kind of training and sharing of the information can be useful for the users to protect the network and all access points that could reduce the issues related to cybercrime with their accounts. Another cost-effective approach to protecting the data is the implementation of multi-factor authorisation (Lamba et al., 2017). This will involve extra security steps for users to login into their account onto the network. This will prevent unauthorised access and help to improve the security level. If the users follow the MFA tool for authentication can be helpful to improve the network security and managing the tasks according to the policies of the university.

The end-users security software is providing the facility for scanning the computer for malicious codes and support to remove them from the computers. According to Pandey, and Misra, (2016), the Master Boot Record (MBR) technique is useful for the encryption of the data and hard drives and detecting the threats. The implementation of electronic security encryption is helpful for the real-time detection of the malware issues through analysis of the heuristic and behavioural analysis of the program or code. The proper monitoring of the network and devices can be a good approach for protecting from the cyber issues and improves the security. The knowledge of the potential behaviour of the programs and devices is useful for identifying the issues that could lead to cybercrime.

## 2.5 Increasing the complexity of Cybercrime in the education sector

As stated above, universities contain a large amount of data regarding students' personal information. The following research studies are the top cybersecurity factors that businesses or the education sector today must consider (Syed, Ahmad, Alaraifi, & Rafi, 2020)

- Increasing complexity, frequency, and scale of cybercrimes
- Leakage of sensitive data, malicious or inadvertent
- Loss of intellectual property
- Strengthening of regulations
- Interconnection of company networks and process control networks

Employees' cybersecurity awareness is an essential factor for preventing and securing organisations from cyber threats and to be aware of security threats while using the Internet (Li, Xu, He, Chen, & Chen, 2016). Employee's usage of the Internet has become an integral part of any organisation's everyday operations. Technologies also provide opportunities to exploit new markets and respond to the specific needs of clients. University employees' understanding of cybersecurity awareness is essential considering their daily use of the Internet. Since the beginning of the technology revolution, higher education institutions start following technological innovation and digital transformation techniques (Syed et al., 2020).

Universities face several issues in changing the traditional education system to an e-learning system. Operational risk is one of the challenges faced in this transformation process. There are some policies and tips for protecting from the cyber issues that involve timely update the software and operating system of the device and the network. This will help in implementing the latest security patches for improving the security level. In addition to this, the utilization of the anti-virus software will be helpful for the management to identify and remove the potential threats from the computer that could harm the data and files. The trick and knowledge related to not opening the attachments from the unknown senders and links can be useful for protecting the devices from infected malware (Walker-Roberts et al., 2020). Moreover, university management needs to set strong passwords and validation codes for restricting unauthorised access.

**2.6 Prevention measures from cyber attack**

The universities need to craft improvement in the measures and policies to protect the data from the cyber attack. These organizations are facing various issues and challenges that are affecting the reputation and trust of the stakeholders which directly affect the financial gain of the universities. According to Puthal et al., (2017), there are different types of approaches available for protecting the data and computer from the cyber-attack or other digital crimes. The foremost approach for protecting the computer from the cyberbullying of attack is providing the knowledge and information related to security principles of the management or other users of the network. This is an economical and simple step for improving cybersecurity. In addition to this, the use of firewalls for the internet connection can be useful to increase the level of security and overcome the issues related to data theft. This is a major issue for the universities to maintain the secure network connection and check the IP addresses of users to identify the unauthorized access into the portal. Apart from this, it is necessary to secure the Wi-Fi network for improving the safety as it will limit the access of the network and control the unauthorized access into the network. This could be useful for the universities to manage the security operation and identifies the potential threats by optimizing the security checking. The proper monitoring and standers approaches for login and access of the university data can help to craft improvement in the networking and offering the online education and information to the students and staff members.

## 3. Research methodology

For conducting the research successfully and in the right manner, the researcher needs to determine the method of the research. The selection of the right method is beneficial for achieving the goals and objectives. To choose the right method for the research, the consideration of three aspects is essential such as divers, barriers and segmentation. The method of research involves the quantitative and qualitative and mix type. Qualitative method of research is applied for collecting the data that influence the opinion of the people to examine and explaining the facts considering the aim and objectives of the research. This method is allowing the researcher to collect and analyse the secondary data from the authentic sources and critically analyse the different aspects of the topic of the study (Mohajan, 2018). The quantitative method of research is applied to discussing the hypothesis derived from the theories. This kind of method is beneficial for the analysis of the objectives of the study. Apart from this, the mixed method of research is used for overcoming the drawbacks of both qualitative and quantitative methods of research and involve theoretical and numerical data to improve the validity and reliability of the research and meet the objectives more professional manner (Basias, and Pollalis, 2018).

For the current research, the researcher has applied the qualitative method for collecting the data and analysis. According to this method, the researcher has collected the non-numerical data from the websites of the universities of Saudi Arabia to critically analyse the policies related to cybersecurity. This has helped to gain the information related to management and cybersecurity concern and approach of the universities of Saudi Arabia (See Appendix A for the list of the universities). Apart from this, the researcher has observed the policies of the universities to get knowledge of the measures implemented for protecting the data and piracy of the management.

**3.1 Research Approach**

To conduct the research, there are two types of research approaches that have been used that involve the inductive and deductive types. The inductive approach of research is used for developing theories considering data analysis and observation. Apart from this, the deductive approach is used for proving the theory by making emphasis on the aim and objectives of the research. This kind of approach is useful for discovering new phenomena considering the findings of the previous studies and analysing the different perspectives that could help in supporting the arguments of the current research. This is a good technique for identifying the gaps in the research and maintaining the focus on the current standards. However, the indicative research is more open and helping to get the knowledge and understanding of the existing situation but having less concern over the past studies for developing the theory (Snyder, 2019).

For the current research, the researcher has applied the deductive approach for analyzing the cybersecurity management and policies of the Saudi Arabian universities. The kind of approach has allowed the researcher to collect the primary data from the websites of the universities and compare the policies for managing cybersecurity. This kind of approach has also helped to develop a valid and reliable conclusion by using the data available through the authentic sources and explaining the concept and variables that have influence the findings of the study. The consideration of existing data and information has helped in proving the theory and satisfying the aim and objectives of the study to meet the desired outcome to analyze the issues that affecting the networking, data interchange and offering the information to management and students using the online portal of study in a secure manner.

**3.2 Research Design**

Research design is an important element for managing and systematically completing the study. This section of the study is providing the detail information related to the what questions need to be answered and when will be the study carried out as well as what type of data is going to be used for conducting the study. In addition to this, the research design is helping the researcher to select the technique for collecting and analysing the data according to the type of study. There are three types of designs have been used for managing the study that involves the descriptive, exploratory and casual. The descriptive design of the study is used for collecting and analysing the information related to large and specified groups by focusing on the two variables of the study. The exploratory design is used for analysing the issues by making emphasis on the objectives of the study. This design is flexible and supports the researcher to explore the new ideas and plan the actions that help in meeting the aim of the study. Moreover, it is providing insight into more subjective matters that influence current policies and standards of the chosen topic (Wiek, and Lang, 2016). Apart from this, the casual design of the research is applied for analyzing the cause-effect relationship of the study and determining the flow of the data collection approach using the various techniques.

For managing the current research for analysing the cybersecurity management and policies of the Saudi Arabian universities, the researcher has selected the exploratory research design. This design helps complete the qualitative research and gain an understanding of the issues and approaches a more professional manner. By using this design, the researcher has done the strategic planning for maintaining the research in a systematic manner that has saved the time for collecting the data from the websites of the universities. The utilization of flexible sources for completing the research has also helped to invalidate the data and increasing the reliability as data is collected from the unbiased sources. Moreover, the researcher has involved the users and included the internal reports of the universities based on cybersecurity and threats to improving the authenticity of the study. This kind of design and approach has helped the researcher to conclude the study simply and more easily by focusing on the aim and objectives of the study.

**3.3 Data Collection**

Data collection is an important aspect of the study that driver the whole research and influences the overall actions and outcome of the research. To increase the validity and reliability of the research, the researcher needs to collect the authentic data and support the aim and objectives. However, this research collected the data from the websites of the Saudi Arabian universities to understand the policies and measures that applied by the management for overcoming the threats of the data hacking, phishing and unauthorised access into the portal. Apart from this, the consideration of information provided in the literature review has also helped in supporting the arguments and discussion of the finding. This kind of data collection approach of the researcher has also supported in minimizing the cost of the study and completing it in the estimated time.

**3.4 Sampling and data analysis**

To conduct the primary research, the researcher needs to define the sampling method and size of the sample. The prior identification and selection of the population and characteristics of the respondents are helpful for the researcher to develop a logical conclusion from the finding of the data analysis. For the sampling, there are two types of methods have been used that involve the probability and non-probability sampling. The probability sampling method is applied when the researcher knows the entire population. Apart from this, the non-probability methods are applied where the researcher is free for choosing the sample according to own convenience (Humphries, 2017). Now, to complete the study about analysing the cybersecurity management and policies of Saudi Arabian universities, the researcher has applied the non-probability method and selected the sample size of 10 universities of Saudi Arabia.

Data analysis is the most important part of the research that helps in developing the valid conclusion and meets the objectives and aim of the study. The current study is based on the qualitative type and exploratory research design. Therefore, the researcher has chosen the thematic and frequency distribution analysis method for the analysis of the data collected through the primary method. The researcher has developed the themes based on the questions and evaluated the collected data. To maintain the systematic approach for the presentation of the data, the researcher has developed the graphs with an interpretation of the respondents related to the challenges and policies of the universities. This kind of presentation will help the readers to understand the response and approach of the analysis that used for evaluation of the data (Ulmer, 2017). Moreover, this kind of approach is beneficial for maintaining transparency and offering the recommendations to maintain the effectiveness in the cybersecurity for the universities of Saudi Arabia. Apart from this, the researcher has provided a discussion over the findings of the analysis to satisfy the aim and objectives of the research.

**4 Data Analysis**

The important part of the research is the data analysis which provide the right information related to the study and help the readers to understand and well knowledge the process of the researcher evaluate the sample to meet the aim and objectives of the research. The researcher has developed different theme by focusing on the approaches of the selected universities and analysed of their websites and students portals. Furthermore, the researcher has observed the actions and approaches of the students and staff to get the access of the information using the websites of the universities. This can help to understand the related actions and issues of the universities as well as own experience by analysing through the cyber policies and standards of the universities.

*Theme 1: Experience of cybersecurity at university*

There are many universities in Saudi Arabia that offering different types of courses and classes according to global standards and providing support to the students through the website. The management has developed a dynamic website that offers the information related to course, library and future activities in the time of Covid-19. According to the analysis of the 10 universities websites from Saudi Arabia, the leading universities like King Abdulaziz, Jazan University, King Faisal University and Qassim University have updated the privacy policies for the cybersecurity and protection of the data. The managements have developed the sites that have protection for unauthorized access and seeking for the registration before login into account of the university. The knowledge is the legal framework and follow us with the regulations for the cybersecurity guidelines is supporting the universities to maintain the quality network and provide the safety for the users of the website. Apart from this, the King Fahd University of petroleum and minerals, University of Hail and Taibah University are not secured according to the guidelines of the IT to protect the content and information of the visitors of the sites and students. The changes in the current policies and guidelines for the cybersecurity are not updated in the websites if the universities and having a negative impact on the privacy and leading towards the issues like hacking and phishing. Now, the management of the universities need to understand the challenges and get the support from the IT companies to increase the security level and maintain the good practices for storing and access of the accounts of the staff and students.

***Theme 2: Major issues observed***

According to the analysis of the online portal of the universities of Saudi Arabia, it can be considered that some of the online portals are having good practices related to maintaining cybersecurity. The universities such as Jazan University, Qassim University, King Abdulaziz and King Khalid University have implemented the protocols and applied tools that are checking the IP addresses and approach of the users to maintain the security.

The issues that identified in the online education process involve the phishing of the data, data leakage and IoT ransomware. These are having a direct impact on the devices that are connected with the network of the organization. The users are facing the challenges related to the stealing of personal information from their accounts and changes in the bio and other data without their permission. As per the analysis of the websites of the universities, the lack of monitoring and regular updating of the security versions the users are facing issues. Most of the issues have been found in King Fahd University of petroleum and minerals, University of Hail and Taibah University. The users have a complaint about these but lack of concern and response from the management is influencing the issues related to cybersecurity.

The phishing of information about the passwords, usernames and payment details for the course is having a negative impact on the process of the data using the online portals of the universities. The users are not trusting the entities and not providing the personal details through contact services of the websites of the universities. The lack of security measures is having a direct impact on the reputation of the educational institutions and raising the questions for the management to maintain the standards. Moreover, the evaluation of websites related to cyber policies have highlighted the issues related to data privacy as there are no policies for cloud-based data storage and access to it. The universities are not following the General Data Protection Regulation that is defined by the IT authorities of the nation. The leakage of information is a major threat for the students and staff members as it can be used for personal benefits or fraud. The threat of stealing personal data is requiring strict actions from the universities to overcome the threats related to cybersecurity. In addition to this, the Distributed Denial of Services (DDoS) attacks are the most common type of cyber issues in ICT that affecting the level of education venue.

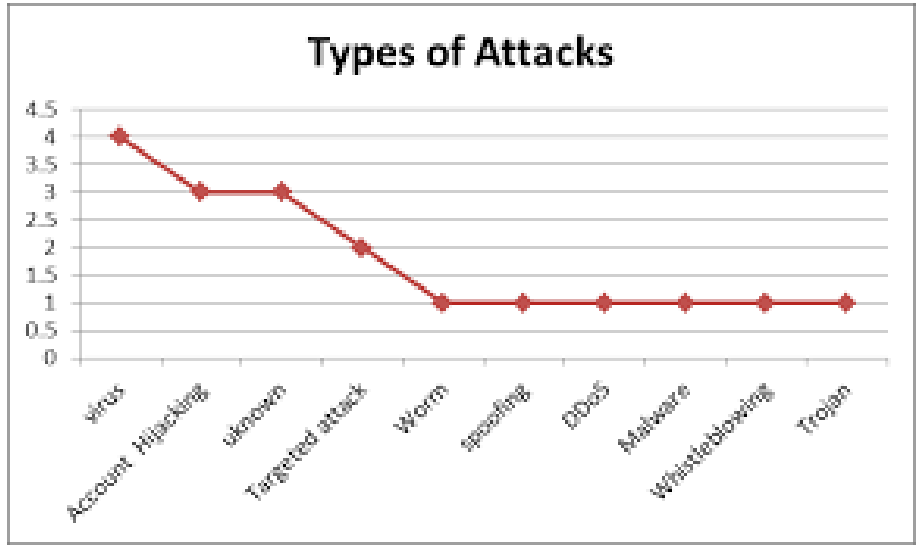

Figure 1: Cyber issues

*Theme 3: Awareness of the policies related to cybersecurity*

According to the evaluation of the websites of Saudi Arabian universities, it has carried out that only 3 out of every 10 students from different universities are having the knowledge of the policies related to cybersecurity. Most of the network users are not aware of the policies and regulations that need to be considered for maintaining the protection of the data and information they are sharing through the network. Apart from this, the universities management is not offering any training and information related to the cyber policies that might help in increasing the safety of the data and support in protecting the data. The IT facilities are operated by approval of the university management.

According to the analysis of websites of the universities, there are some policies and privacy-related regulations are mentioned that are helping the users to understand the approach for using the network. The Jazan University, Qassim University, King Abdulaziz and King Khalid University have implemented the protocols and applied tools and offering the information related to the utilization of the cloud services and use of the protocols for managing the details of the individual. The universities are also offering information for the responsibilities of the individual for using the network and login to the portal. The policies are applied for users and devices of the IT facilities and services and the IT team is configuring the computers and laptops that helping to protect the information and data from hacking and phishing. However, it is the responsibilities of people to get the knowledge and information of the policies and legal obligations to maintain their privacy and protection from the issues related to cybercrime. The management and students have to get the knowledge and understanding of the applicable statutory legislation for maintaining the standards and following the guidelines of the universities. In addition to this, the universities have to provide the information for attack and cyber issues that might cause problems for the individual through advertising and IT support.

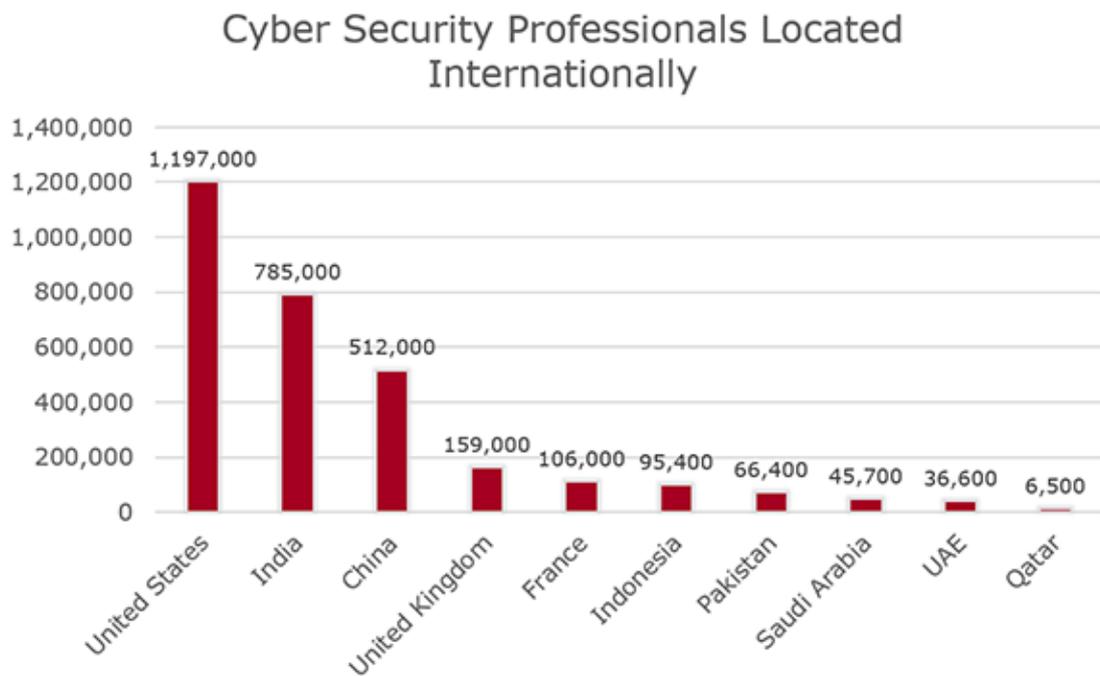

Figure 2: Awareness and experts

***Theme 4: Awareness of the policies related to cybersecurity***

There are various sources that are offering information and knowledge about the cybersecurity approaches and framework for improving the protection level for the data. To gain the knowledge of cyber policies, the literature review and analysis of website have provided the name of sources such as online articles, discussion with the IT experts, focusing on the privacy policies mentioned on the website of the universities and joining the workshop. All the users of university are not from the technical background and not having much understanding of the issues and cybersecurity. To gain knowledge after facing the issues, the users of the IoT and ICT services of universities are giving preference to reading the online articles and approaches to utilizing the particular software. This kind of approach has helped them to understand the operational process and developing the technical skills that might support in mitigating the vulnerabilities.

The implementation of cyber securities is not having a direct impact on the approach of an individual as hackers are using the advance tools and techniques for phishing and stealing the data. According to the analysis of the policies and framework if university's cybersecurity, it can be considered that major sources of gaining the knowledge of the policies and cybersecurity measures are online reading and follow up of the standards mentioned on the website of the university. The major legislations of the Saudi Arabian government for cybersecurity involve the Telecom Act 2001, Anti-Cyber Crime Law 2009 and Electronic Transaction Law 2007. The management and students of the universities can get information about cybersecurity and approaches by reading these articles and improve practices for controlling, regulating and applying the legal framework for managing the online accounts.

According to the analysis of the websites of universities, it can be considered that Jazan University, Qassim University, King Abdulaziz and King Khalid University, King Fahd University of petroleum and minerals, University of Hail and Taibah University have mentioned the policies and framework for the protection of the data of individual by following the policies. The websites of these universities are offering the knowledge and understanding about the legal information related to security and offering of the personal information to authorized links of the university. The management has clearly stated that the use of privacy control measures as per the government and university approaches for managing cybersecurity is helpful for overcoming the risks.

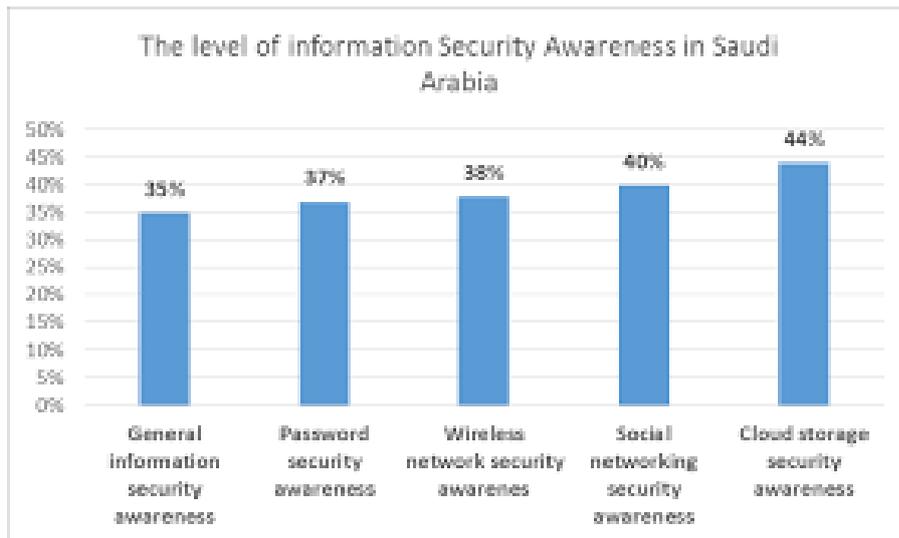

Figure 3: Knowledge

Apart from this, the universities are getting the knowledge and understanding from the leading cyber companies for protecting their network, operating system and cloud data. As per the analysis, the major universities of Saudi Arabia are taking assistance from the United States' National Institute of Standards and Technology (NIST) for gaining the knowledge about the cybersecurity functions and improving the policies for the internet security. In addition to this, the USA organization is offering the knowledge and creating the awareness about the cloud computing that involves the identification of safe and private network for offering the learning and education to the students and providing the facility for the controlling the activities and managing the data safe and secure. Moreover, the organizations have to develop the IT expert desk for offering the information related to the policies and approach for the individual to manage the cybersecurity operations. The comprehensive knowledge and information related to security policies will help to overcome the issues. The lacking in monitoring and understanding of the policies are creating threats for Saudi Arabian universities to manage the cyber issues and improve security.

*Theme 5: Changes required for improving the cybersecurity*

The major changes that suggested by the respondents involve continues development and improvement in cyber defences, monitoring and review of the practices and activities of people who involve in the unauthorized access of the data. Moreover, it is important for the universities to develop the risk audit and compliance committees that are responsible for the implementation of the practices and offering of the information to the management and users. The analysis of sites and cybersecurity policies of universities has suggested that overseeing of the cybersecurity and capability control is also essential for improving the approach of the universities for managing the IT facilities. The framework of universities is lacking and not meeting the standards of operational support facilities. The programmers and testing team is required for supervising and identifying the gaps in the security services and implementation of the support plan to fix it. According to the principles of the data security, the changes that required for mitigating the issues of Saudi Arabian universities involve the adoption, compliance and regulatory management.

There is the requirement of cloud security coordinator centre in Saudi Arabia that will look after the equipment and provide the training to the IT experts for creating the awareness about the issues and approaches that help in protecting from cybercrime. As per the analysis, the major requirement for strict implementation and adherence to all current cybersecurity laws in the universities and compulsory training for the management for better understanding of the cloud data and protection.

## 5 Discussion and conclusion

The research has analysed the cybersecurity management and policy issues in Saudi Arabian Universities and identified that there are various issues that are influencing the approach of the staff and students to maintain the privacy and other security measures for data. According to analysis, the university is lacking in infrastructure development and continues improvement in the cyber policies and regulations. The students and management are also not aware of the policies and protection standards that required for improving the data security to maintain the safety of their personal information and digital learning approaches. The research has also identified the computing technologies used by the universities for offering education and securing there data and activities over the network. In addition to this, the study has analyzed that the universities have developed websites that offering information related to the course and providing the study material. The students can login into the portal through personal email id.

In the current scenario, the higher educational institutions are using the digital technologies and devices for offering the learning and sharing of the study material with the students. This is helping to connect with the students and monitor their learning more effective manner. Moreover, this type of learning is also helping the students to get real-time support from the website of the universities by accessing the library by creating the account. However, the cybersecurity issues related to phishing, hacking attack of the various viruses and hacking is influencing the approaches of the individual. There is no awareness about the issues and policies for maintaining the standards of the security which influencing the management approaches and implementation of good practices (Cheung et al., 2011). The knowledge and application of the global policies and regulations for cybersecurity can be useful for the university managements to overcome the impact of the hacking, malware attack, phishing of the data and misuse of the cloud storage. The issues with ICT related to security are involving the privacy means, lack of understanding of the utilization of the tools and applications for connecting with the portal of the university. Moreover, information and knowledge of the multi-layer cybersecurity approach is the best way to protect the data and systematically manage the information.

According to the analysis of the study, the universities of Saudi Arabia are looking to update the cybersecurity measures and policies to increase the standards of cyber activities and actions. As per the analysis, the implementation of the new technique is involving end-to-end protection. The implementation of electronic security encryption is helping for the real-time detection of the malware issues through analysis of the heuristic and behavioural analysis of the program or code. Apart from this, the study has identified that universities of Saudi Arabia are using different software and suits that are helping to increase the level of cybersecurity. In addition to this, the implementation of cloud-based security protection is helping the universities and management to secure the browsing and protecting from the virus attack as it blocked the site of links.

In addition to this, the lack of implementation of intrusion policies for detecting unauthorized access is also affecting the cybersecurity of the universities. The study has discussed and provided the information related to the implementation of exceptions policy and host integration that can be useful for the universities to protect the data and maintaining the security level. The exceptions policy provides the ability to exclude applications and processes from the detection of the virus and scan the data. In addition to this, host integration is helping the users and network providers to enforce and restore the security of the client computer that influence the level of the security and give the information about the implementation of protocols of accessing the portal. Apart from this, the research has identified the approaches that required the changes in the policies and framework for managing data security at the university level.

By considering the analysis and identification of cybersecurity issues and management approaches of the Saudi Arabian universities, it can be considered that the identified 10 universities are facing different issues. The management has to look into the challenges and have to plan the development for the betterment of the secure network. The findings can be seen as a reinforcement of developing a software application artefact for information security officers. Any of the design science approaches (e.g. Miah, Gammack & McKay, 2019; Miah, Gammack & Kerr, 2012; Miah, Kerr, Gammack & Cowan, 2008) could be adopted for developing and evaluating a new purposeful security solution artefact (e.g. with enhanced accessibility (Miah, 2004) and advanced features (Miah, 2008; 2009; Miah & Ahamed, 2011; Miah & Gammack, 2008). Also we produce some recommendations that could be useful for the universities to manage the issues and craft improvement in the cyber policies and implement them strictly:

- The universities can take support from the international agencies that are identifying the issues and developing the tools for protecting the data. The American companies are best in this business and offering the best software and framework that could help the universities to improve the standards. The universities can contact with the NIST of USA for constancy and offering the support for continuous improvement in the policies and infrastructure of the organizations

- For protecting from the cyber issues and craft improvement in the services options and data management, it is recommended to the university management to conduct the regular monitoring of the servers, devices, network and approaches of the users. The proper auditing will be helpful for the management to understand the factors that are creating barriers and make real-time efforts to improve the services standards.

- The management of universities has to develop the IT desk for protecting cyber issues. This kind of approach will help in influencing the security measures. The development of the IT expert desk for offering the information related to the policies and approach for the individual to manage the cybersecurity operations. The comprehensive knowledge and information related to security policies will help to overcome the issues. In addition to this, the use of AI tools and technology will be helpful for improving the protection which will monitor the approach of the users and functioning of the operating system.

**Appendix A**
**List of Universities**
1. King Abdulaziz University
   https://www.kau.edu.sa/Home.aspx
2. King fahd university of petroleum and minerals
   http://www.kfupm.edu.sa/ar/Default.aspx
3. King Faisal University
   https://www.kfu.edu.sa/sites/Home/
4. King Khalid University
   https://www.kku.edu.sa/
5. King Saud University
   https://www.ksu.edu.sa/en/
6. Qassim University
   https://www.qu.edu.sa/
7. Taibah University
   https://www.taibahu.edu.sa/Pages/AR/Home.aspx
8. Taif University
   https://www.tu.edu.sa/
9. University of Hail
   http://www.uoh.edu.sa/Pages/default.aspx
10. Jazan University
    https://www.jazanu.edu.sa/